\pdfoutput=1
\documentclass[preprint,12pt]{elsarticle}

\usepackage[margin=1in]{geometry}

\usepackage[hyphens]{url}
\biboptions{sort&compress, square, comma}
\usepackage[breaklinks=true, linkcolor=blue, citecolor=blue, colorlinks=true]{hyperref}

\usepackage{graphicx}
\usepackage{caption}
\usepackage{subcaption}

\usepackage[version=3]{mhchem} % Formula subscripts using \ce{}, e.g., \ce{H2SO4}
\usepackage{bm}
\usepackage{latexsym, amsmath, amssymb}

\usepackage{mathtools}
\usepackage{exscale, relsize}
\usepackage[retainorgcmds]{IEEEtrantools}
\usepackage{nicefrac}

\usepackage{booktabs,longtable,multicol,multirow}

\usepackage{cancel}
\usepackage[super]{nth}

\usepackage[T1]{fontenc}
\usepackage[english]{babel}
\usepackage{csquotes}
\usepackage{textcomp}
\usepackage{subfloat}

\usepackage{siunitx}
\DeclareSIUnit\atm{atm}

\graphicspath{{./figures/}}

\usepackage[textsize=small,textwidth=2.5cm]{todonotes}

\journal{Fuel}

\begin{document}
\begin{frontmatter}

\title{Predicting fuel research octane number using Fourier-transform infrared absorption spectra of neat hydrocarbons}

\author[osu]{Shane~R.\ Daly}
%\ead{dalys@onid.orst.edu}

\author[osu]{Kyle~E.\ Niemeyer}
%\ead{Kyle.Niemeyer@oregonstate.edu}

\author[chev]{William~J.\ Cannella}
%\ead{bijc@chevron.com}

\author[casc]{Christopher~L.\ Hagen\corref{cor1}}
\ead{chris.hagen@oregonstate.edu}

% addresses
\address[osu]{School of Mechanical, Industrial, and Manufacturing Engineering\\
	Oregon State University, Corvallis, OR 97331, USA}
\address[chev]{Chevron Energy Technology Company\\
	Richmond, CA, 94802, USA}
\address[casc]{School of Mechanical, Industrial, and Manufacturing Engineering\\
	Oregon State University--Cascades, Bend, OR 97701, USA}

\cortext[cor1]{Corresponding author}

%%%%%%%%%%%%%%%%%%%%%%%%%%%%%%%%%%%%%%%%%%%%%%%%%%%%%%%%%%%%%%%%%%%%%%
\begin{abstract}
% One or two sentence brief introduction to field
% Two or three sentences of detailed background
% One sentence clearly stating general problem
% One sentence summarizing main result ("here we show")
% Two or three sentences explaining what the main result reveals in direct comparison to what was thought to be the case previously, or how the main result adds to previous knowledge
% One or two sentences to put the results into a more general context
% Two or three sentences to provide a broader perspective

%Liquid transportation fuels are characterized by performance metrics such as %octane rating determined by costly and time-consuming physical tests.
%These tests can be avoided through the use of multivariate statistical models %built using the optical spectra and known behavior of characterized fuels, as %long as fuel sale restrictions (mandatory use of CFR) do not apply.
%Existing techniques inform these models using information about existing, %similar fuels---for example, training a model for gasoline RON with a large %number of characterized gasoline samples.
%While this yields the most accurate predictive models for these fuels, this %approach lacks the ability to predict characteristics of fuels outside the %training data set.
Liquid transportation fuels require costly and time-consuming tests to characterize metrics, such as Research Octane Number (RON) for gasoline.
If fuel sale restrictions requiring use of standard Cooperative Fuel Research (CFR) testing procedures do not apply, these tests may be avoided by using multivariate statistical models to predict RON and other quantities.
Existing techniques inform these models using information about existing, similar fuels---for example, training a model for gasoline RON with a large number of characterized gasoline samples.
While this yields the most accurate predictive models for these fuels, this approach lacks the ability to predict characteristics of fuels outside the training data set.
Here we show that an accurate statistical model for the RON of gasoline and gasoline-like fuels can be constructed by ensuring the representation of key functional groups in the spectroscopic data set are used to train the model.
We found that a principal component regression model for RON based on IR absorbance and informed using neat and 134 mixtures of \emph{n}-heptane, isooctane, toluene, ethanol, methylcyclohexane, and 1-hexene could predict RON for the 10 Coordinating Research Council (CRC) Fuels for Advanced Combustion Engine (FACE) gasolines and 12 FACE gasoline blends with ethanol within 34.8\textpm 36.1 on average and 51.2 in the worst case.
We next studied the effect of adding 28 additional minor components found in the FACE gasolines to the statistical model, and determined that it was necessary to add additional representatives of the branched alkane and aromatics classes to reduce model error.
For example, adding 2,3-dimethylpentane and xylene to the previous model allowed it to predict RON for the 22 target fuels within 0.3\textpm 4.4 on average and 7.9 in the worst case.
However, we determined that the specific choice of fuel in those classes mattered less than ensuring the representation of the relevant functional group.
This work builds upon previous efforts by creating models informed by neat and surrogate fuels---rather than complex real fuels---that could predict the performance of complex unknown fuels.

\end{abstract}


\end{frontmatter}

%\linenumbers
%\renewcommand\linenumberfont{\normalfont\tiny}

%%%%%%%%%%%%%%%%%%%%%%%%%%%%%%%%%%%%%%%%%%%%%%%%%%%%%%%%%%%%%%%%%%%%%%%%%%%%%%%%%%%%%%%%%%%%%%%%%%%%
\section{Introduction}
\label{sec:intro}
%%%%%%%%%%%%%%%%%%%%%%%%%%%%%%%%%%%%%%%%%%%%%%%%%%%%%%%%%%%%%%%%%%%%%%%%%%%%%%%%%%%%%%%%%%%%%%%%%%%%

Research octane number (RON), determined by ASTM-CFR standard testing procedure D2699-15~\cite{ASTM-D2699-15}, indicates a fuels' resistance to autoignition under specific engine operating conditions.
RON and the other ASTM 4814 fuel specifications dictate several attributes necessary to operate in the installed fleet of vehicles.
In 2014, 136.78 billion gallons of gasoline were consumed in the US~\cite{EIA:2015}---all of which need to meet those quality specifications.
Determining the RON of fuels using a Cooperative Fuels Research (CFR) engine costs over \$200,000 for the capital investment (among considerable lab modifications) takes 20 minutes, and also requires trained technicians\slash operators.     

In an effort to reduce this testing burden, researchers sought out more cost-effective and faster noninvasive optical techniques for determining RON, among other fuel specifications, by way of statistical analysis.  Vibrational spectroscopy, such as infrared absorption (IR) and Raman spectroscopy, has proved to be a reliable method for fuel characterization.  The work of Kiefer~\cite{Kiefer2015} highlights current technical advances in the context of fuel characterization, overviews fundamental theory, and discusses advantages/ disadvantages of the various techniques currently in use today.  Now, a brief sequential overview of literature utilizing vibrational spectroscopy in conjunction with statistical analysis will be discussed.    

Kelly et al.~\cite{Kelly:1989ui} determined 10 ASTM specifications including RON, Motor Octane Number (MON), vapor pressure, specific gravity, bromine number, and contents of aromatic, alkene, saturate, sulfur, and lead using a short wavelength near infrared (SW-NIR) scanning spectrophotometer (\SIrange{660}{1215}{\nano\meter}) and multivariate analysis to correlate the spectra to the performance metrics.
For example, the group showed that RON of gasolines can be predicted to a standard error within 0.4--0.5~\cite{Kelly:1989ui}, which is better than the ASTM RON test itself at \textpm 0.7~\cite{ASTM-D2699-15}.
The original work of Kelly et al.~\cite{Kelly:1989ui} inspired other investigations to enhance their technique, consider alternate fuels, or to predict other fuel performance metrics.
To briefly touch on these alternate studies, Williams et al.~\cite{Williams:1990aa} instead leveraged FT-Raman spectra (\SIrange{3200}{600}{\centi\meter^{-1}}) to predict cetane index and cetane number to \textpm 1.22 and 2.19, respectively.
Cooper et al.~\cite{Cooper:1995aa} applied a similar methodology as Williams et al.\ (using Raman spectra at wavenumber ranges of  \SIrange{2510}{3278}{\centi\meter^{-1}} and \SIrange{196}{1851}{\centi\meter^{-1}}) to predict MON, RON, and pump octane number to within \textpm 0.415, 0.535, and 0.410, respectively.
Litani-Barzilai et al.~\cite{LitaniBarzilai:1997wp} combined near-IR (\SIrange{700}{1000}{\nano\meter}) and laser-induced fluorescence (\SIrange{250}{500}{\nano\meter} third and fourth harmonic) spectra to predict 10 physical specifications; e.g., RON and MON were predicted to within \textpm 0.33 and 0.27, respectively.
The more recent work of Kardamakis and Pasadakis~\cite{Kardamakis:2010db} presents an efficient multivariate analysis technique that predicts RON within \textpm 0.26 using a limited data set in comparison to previous studies; this work also provides a succinct history of efforts in this field.
There are many additional studies to the short list previously mentioned that consider various optical and multivariate analysis techniques to predict performance parameters of fuels~\cite{Swarin:1991aa,Choquette:1996aa,Fodor:1996aa,Korolev:2000tb,Balabin:2008aa,Monteiro:2009kk,Morris:2009cg,Veras:2010jt,Tomren:2012aa}.

Various commercial devices utilize these principles to rapidly predict relevant properties of gasoline and diesel fuels based on optical characteristics.
For example, the Zeltex ZX 101C octane analyzer~\cite{Zeltex} passes radiation from light emitting diodes through optical filters and gasoline samples (14 static wavelengths ranging from \SIrange{893}{1045}{\nano\meter}).
The light is collected on a photodetector and processed for absorbance at the wavelengths of interest, with a total measurement time of \SI{20}{\second} and accuracy of \textpm 0.5 RON units~\cite{Zeltex}.
The IROX Miniscan IRXpert gasoline\slash diesel analyzer takes a similar approach based on FTIR spectroscopy, collecting a broad absorption spectrum and generating information at \num{12900} wavelengths.
This allows the prediction of 16 total ASTM specifications, and predicts RON with an accuracy of \textpm 0.5 within \SI{80}{\second}~\cite{IROX}. 
This equipment costs less than half of a CFR engine and does not require expert technicians\slash operators.

All the previous approaches using multivariate analysis to predict fuel attributes~\cite{Kelly:1989ui,Williams:1990aa,Cooper:1995aa,LitaniBarzilai:1997wp,Korolev:2000tb,Kardamakis:2010db,Swarin:1991aa,Choquette:1996aa,Fodor:1996aa,Balabin:2008aa,Monteiro:2009kk,Morris:2009cg,Veras:2010jt,Tomren:2012aa} used existing, real-world fuel samples (i.e., existing gasoline, diesel, jet fuels) as the training data set to predict performance attributes of those specific fuels.
This work used hydrocarbons---neat or combined as mixtures for gasoline surrogate fuels including up to five neat components---to provide model input for predicting RON of the Fuels for Advanced Combustion Engines (FACE) gasolines designed by the Coordinating Research Council (CRC) and manufactured by ChevronPhillips Chemical Co ~\cite{Cannella:2014aa}.
With this novel approach, a sensitivity analysis can then target neat hydrocarbons and classes (i.e., functional groups) to develop and optimize spectroscopic surrogates for the FACE gasolines.
These spectroscopic surrogates most simply represent the bulk auto-ignition behavior (through statistics) of the FACE gasolines.
Researchers and industry alike can then predict RON for future fuels (e.g., new, alternative, regarding advanced engines) that may otherwise not be accurately represented spectroscopically by traditional fuels used today.
Here, the statistical models created are robust in that they are informed on a fundamental level. This mitigates the issue of creating a model informed by existing fuels that may be physically and spectroscopically different to future fuels---inaccurate prediction of the future fuel would result.

This work uses the FACE gasoline for the fuel and RON to represent the fuel performance parameter. RON is readily obtained for neat hydrocarbons, surrogate and research-grade gasolines, and has previously been shown extensively in literature to correlate well with optical data of quantified gasoline samples.
We test our model by predicting RON for the 10 FACE gasolines and 12 additional blends with ethanol; these represent candidate fuels for advanced internal combustion engines (i.e., future fuels)~\cite{Cannella:2014aa}.

The structure of the paper is as follows.
Section~\ref{S:method} presents the methodology of the approach.
This section includes the neat hydrocarbons and surrogate gasoline mixtures considered in this work, the FTIR spectra collection method, and the development of the statistical model.
Section~\ref{S:results} provides the results and discussion of the predicted RON values of FACE gasoline samples from the developed statistical model.
Lastly, Section~\ref{S:conclusions} summarizes the findings of this study.

%%%%%%%%%%%%%%%%%%%%%%%%%%%%%%%%%%%%%%%%%%%%%%%%%%%%%%%%%%%%%%%%%%%%%%%%%%%%%%%%%%%%%%%%%%%%%%%%%%%%%
\section{Methodology}
\label{S:method}
%%%%%%%%%%%%%%%%%%%%%%%%%%%%%%%%%%%%%%%%%%%%%%%%%%%%%%%%%%%%%%%%%%%%%%%%%%%%%%%%%%%%%%%%%%%%%%%%%%%%%

In the current approach, hydrocarbon components (neat or mixtures of up to five components) informed a statistical model rather than characterized gasoline samples as in prior efforts.
First, the training data set---i.e., the pure hydrocarbon components and mixtures considered to train the statistical model---is discussed.
Second, IR absorbance spectra collection methods and the statistical methodology used in this work are covered.
Lastly, with the statistical model created, the methodology to validate the model is discussed.

%%%%%%%%%%%%%%%%%%%%%%%%%%%%%%%%%%%%%%%%%%%%%%%%%%%
\subsection{Neat hydrocarbons considered}
%%%%%%%%%%%%%%%%%%%%%%%%%%%%%%%%%%%%%%%%%%%%%%%%%%%

Promoted by the literature~\cite{Perez:2012dga,Foong:2014fz,Truedsson:2014ke} as components most relevant to simple fuel surrogates, we primarily considered mixtures of \emph{n}-heptane, isooctane, toluene, ethanol, methylcyclohexane, and 1-hexene.  
These six hydrocarbons will be referred to as the ``primary'' hydrocarbons used in this study.
In brief, the first two components are used to measure RON (also called the primary reference fuels, or PRFs) and represent the straight and branched alkane functional groups, respectively.
Toluene and ethanol represent aromatics and oxygenates, while methylcyclohexane and 1-hexene represent cycloalkane (naphthene) and alkene (olefin) classes, respectively.
This study used the aforementioned neat hydrocarbons in addition to the 134 blends taken from the literature~\cite{Perez:2012dga,Foong:2014fz,Truedsson:2014ke,Truedsson:2014ut}. 
These blends are mixtures of the six hydrocarbons in various combinations ranging from two to five components, primarily consisting of isooctane, \emph{n}-heptane, and a third component; see the supplemental material for the full list.

In addition to the six primary neat hydrocarbons, we also considered hydrocarbons found within the FACE gasolines via detailed hydrocarbon analysis~\cite{Cannella:2014aa}.
Table~\ref{T:pure_fuels} lists these additional 28 pure components, referred to as the ``additional'' hydrocarbons in this work; they will be used to supplement the ``primary'' hydrocarbons.
The hydrocarbon classes of these additional species overlap with the classes from the primary set.
However, an outcome of this study demonstrated that the primary set---common components in gasoline surrogate mixtures~\cite{Perez:2012dga,Foong:2014fz,Truedsson:2014ke}---was not sufficient to physically and spectroscopically represent the FACE gasolines, and species from the additional set were needed (see Section~\ref{S:results}).

\begin{table}[tbp]
\centering
\begin{tabular}{@{}l l l l@{}}
\toprule
Fuel name 		    & Formula       & RON		& Class \\
\midrule
\emph{n}-heptane 	& \ce{nC7H16}   & 0 		& straight alkane \\
isooctane		    & \ce{C8H18}    & 100		& branched alkane \\
toluene			    & \ce{C6H5CH3}  & 113~\cite{Truedsson:2014ke}	& aromatic \\
ethanol				& \ce{C2H5OH}	& 108.5$^{*}$~\cite{Hunwartzen:1989aa,Anderson:2012aa,Foong:2014fz}	& alcohol \\
methylcyclohexane	& \ce{C7H14}	& 74.1~\cite{Perez:2012dga}		& cycloalkane \\
1-hexene			& \ce{C6H12}	& 74.9~\cite{Perez:2012dga}		& alkene \\
2-methylbutane		& \ce{C5H12}    & 92~\cite{Scherzer:1989aa}	& branched alkane \\
2-methylpentane	    & \ce{C6H14}    & 73.4~\cite{mines_db}	& branched alkane \\
3-methylpentane	    & \ce{C6H14}    & 74.5~\cite{mines_db}	& branched alkane \\
2-methylhexane	    & \ce{C7H16}    & 42$^{*}$~\cite{mines_db,Carey:2007aa}	& branched alkane \\
3-methylhexane	    & \ce{C7H16}    & 52$^{*}$~\cite{mines_db,Carey:2007aa}	& branched alkane \\
2,4-dimethylpentane	& \ce{C7H16}    & 83.1$^{*}$~\cite{mines_db,Carey:2007aa}	& branched alkane \\
2,3-dimethylpentane	& \ce{C7H16}    & 91.1$^{*}$~\cite{mines_db,Carey:2007aa}	& branched alkane \\
2,5-dimethylhexane	& \ce{C8H18}    & 55.3$^{*}$~\cite{mines_db,Carey:2007aa}	& branched alkane \\
2,4-dimethylhexane	& \ce{C8H18}    & 65.2$^{*}$~\cite{mines_db,Carey:2007aa}	& branched alkane \\
3-ethyl-2-methylpentane	& \ce{C8H18}    & 87.3$^{*}$~\cite{mines_db,Carey:2007aa}	& branched alkane \\
xylene			        & \ce{C8H10}    & 114~\cite{Owen:1990aa}		& aromatic \\
1,2,3-trimethylbenzene	& \ce{C9H12}    & 100.5~\cite{mines_db}	& aromatic \\
4-ethyl-m-xylene	    & \ce{C10H14}   & 100.6~\cite{mines_db}	& aromatic \\
2-ethyl-p-xylene	    & \ce{C10H14}   & 100.6~\cite{mines_db}	& aromatic \\
1,2,4-trimethylbenzene	& \ce{C9H12}    & 101.4~\cite{mines_db}	& aromatic \\
cumene				    & \ce{C9H12}	& 102.1~\cite{mines_db}	& aromatic \\
1,3,5-trimethylbenzene  & \ce{C9H12}    & 106~\cite{mines_db}	& aromatic \\
1,2,3,4-tetrahydronaphthalene	& \ce{C10H12}   & 96.4~\cite{mines_db}	& aromatic \\
2-propyltoluene		& \ce{C10H14}	& 100.3~\cite{mines_db}	& aromatic \\
1,2,3,4-tetramethylbenzene	    & \ce{C10H14}   & 100.5~\cite{mines_db}	& aromatic \\
cyclopentane		& \ce{C5H10}    & 100.1~\cite{mines_db}	        & cycloalkane \\
cyclohexane			& \ce{C6H12}	& 80.7~\cite{ASTM-STP225-EB}	& cycloalkane \\
butylcyclohexane	& \ce{C10H20}	& 63.8~\cite{Albahri:2003aa}	& cycloalkane \\
1-pentene			& \ce{C5H10}	& 90~\cite{Owen:1990aa}			& alkene \\
2-methyl-2-butene	& \ce{C5H10}	& 97.3~\cite{mines_db}			& alkene \\
2-pentene			& \ce{C5H10}	& 98~\cite{Scherzer:1989aa}		& alkene \\
2-methyl-1-butene	& \ce{C5H10}	& 100.2~\cite{mines_db}			& alkene \\
diisobutylene		& \ce{C8H16}	& 103.8~\cite{Bradley:1997aa}	& alkene \\
\bottomrule
\end{tabular}
\caption{
Pure hydrocarbon species considered in this work. Infrared absorption spectra for all species were acquired via Attenuated Total Reflectance FTIR spectroscopy. $^{*}$ indicates average of multiple values.
}
%\todo[inline]{need to find Colorado school of Mines hydrocarbon database. Also, what are sig figs for the RON values without decimals?}
\label{T:pure_fuels}
\end{table}

%%%%%%%%%%%%%%%%%%%%%%%%%%%%%%%%%%%%%%%%%%%%%
\subsection{IR absorbance spectra collection}
%%%%%%%%%%%%%%%%%%%%%%%%%%%%%%%%%%%%%%%%%%%%%

Absorption spectra were collected using a ThermoFisher Nicolet iS10 FTIR with a single-bounce, Attenuated Total Reflectance (ATR) smart accessory (\SIrange{650}{4500}{\centi\meter^{-1}} at \SI{2}{\centi\meter^{-1}} resolution, crystal type: diamond with ZnSe lens, part number: 222-24700).
A Norm-ject \SI{1}{\milli\liter} latex-free (VWR-53548-001) disposable syringe was used to transfer a few drops of the liquid sample directly onto the ATR crystal.
Prior to spectrum collection of the sample, the FTIR was purged with nitrogen to remove any water vapor contamination.
With the collected light intensity, absorbance was calculated with the following relation (the Beer-Lambert law shown for completeness~\cite{rao2012molecular}): 
\begin{equation}
A(\nu)=\ln \left( \frac{I_o}{I} \right)_{\nu} = \sigma_{\nu} c L \;,
\end{equation}
where $I_{o}$ and $I$ are reference and measured light intensity, respectively, $\sigma_{\nu}$ is molar absorption coefficient (\si{\mole^{-1}\centi\meter\squared}), $c$ is concentration (\si{\mole.\centi\meter^{-3}}), and $L$ is path length of the attenuating medium (\si{\centi\meter}).
No path length or dispersion effects were accounted for in the ATR absorbance results; we found correcting ATR spectra made no difference for the statistical model performance, as discussed next.

\begin{figure}[htbp]
    \centering
    \includegraphics[width=0.75\textwidth]{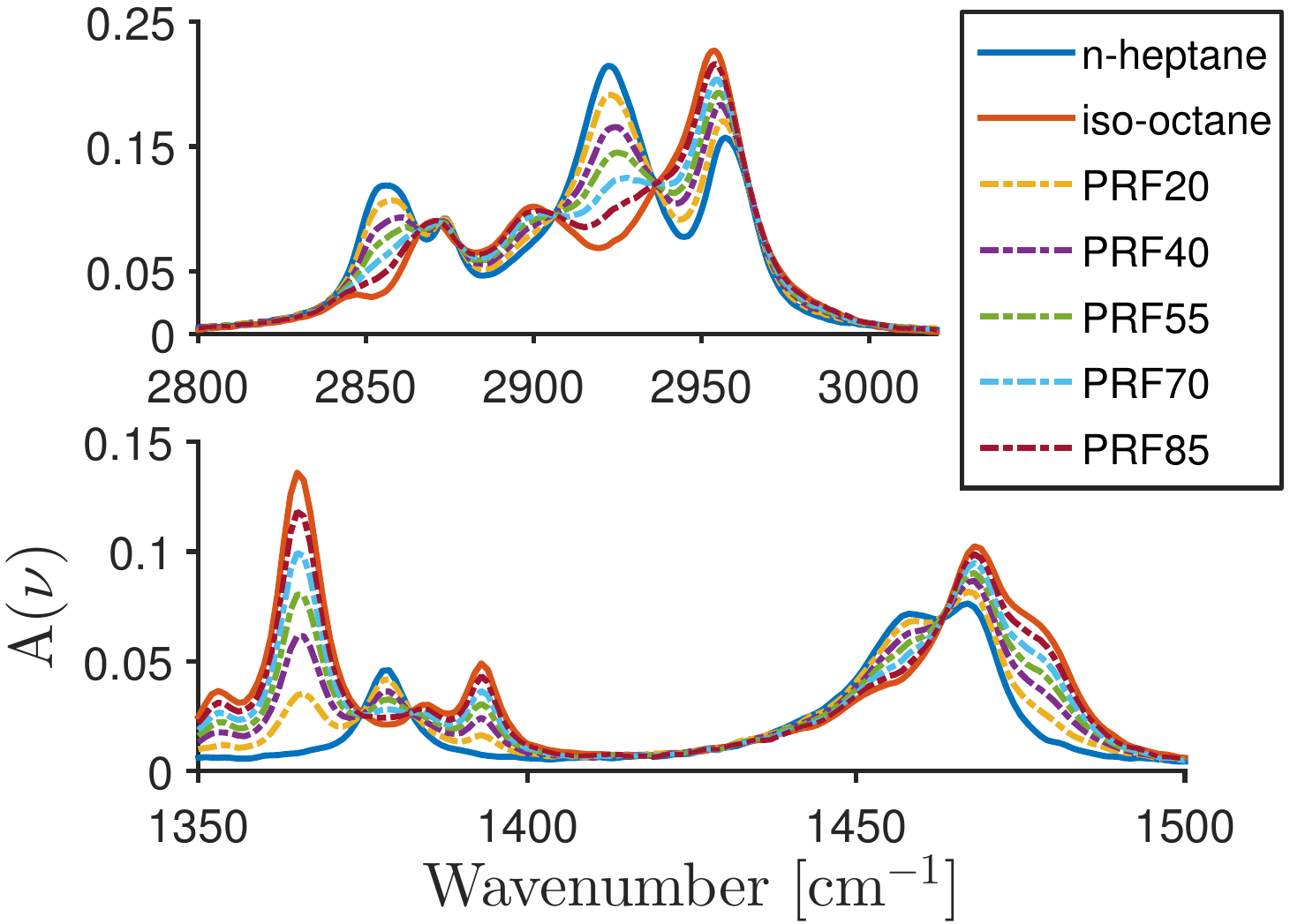}
    \caption{FTIR-ATR absorbance spectra of \emph{n}-heptane, isooctane, and their mixtures (PRFX, where X indicates volume \% of isooctane). The two axes show subsets of the optical frequencies considered where these samples absorb strongly.}
    \label{F:ATR_ex}
\end{figure}

Measurements were also made using a more involved transmission approach by way of a Spectra-Tech EZ-fill\texttrademark{} precision path length optical cell with \SI{3}{\milli\meter} KBr optical slides.
These data were post-processed via the Beer-Lambert law and baseline corrected with the optical constant method guided by literature~\cite{Bertie:1991ge,Bertie:1994aa,Keefe:2002aa,Porter:2009aa,Porter:2009jx}.
We found that when using the same FTIR, no difference resulted in the performance of our approach between using uncorrected ATR (qualitative) or corrected transmission (quantitative) data.
The use of ATR appeared to introduce an instrument function to the data which cancelled when ratioing the incident and transmitted light; as a result, this did not affect the statistical post processing.
Spectra can be collected via ATR an order of magnitude faster than by transmission, in addition to simpler post processing, and thus we chose ATR over transmission-based methods for the current application.
Figure~\ref{F:ATR_ex} shows a representative subset of the collected ATR absorption spectra, for mixtures of \emph{n}-heptane and isooctane.

%%%%%%%%%%%%%%%%%%%%%%%%%%%%%%%%%%%%%%%%%%%%%
\subsection{RON model development}
%%%%%%%%%%%%%%%%%%%%%%%%%%%%%%%%%%%%%%%%%%%%%

\begin{figure}[htbp]
    \centering
    \includegraphics[width=0.75\textwidth]{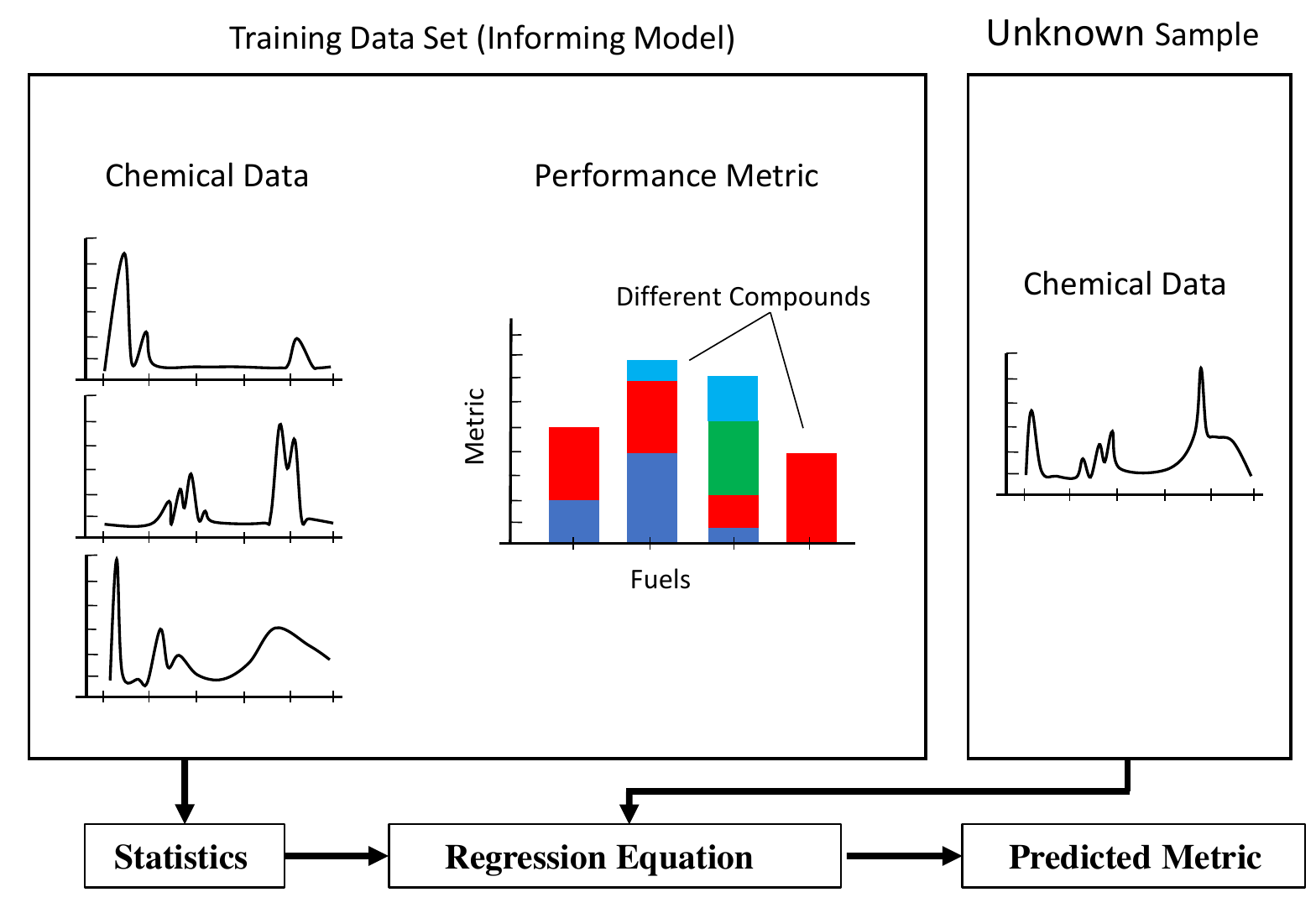}
    \caption{Flowchart depicting the multivariate analysis process.}
    \label{F:flowchart}
\end{figure}

The IR absorbance spectra were correlated to RON by way of principle component regression (PCR). 
PCR identifies patterns in multi-dimensional data and correlates these to an explanatory variable, i.e., a variable that may correlate to the patterns found in the chemical data ~\cite{jolliffe2002principal}.
The authors used PCR as the statistical method, and implemented it via MATLAB software with the built-in function \texttt{pca}.
All principal components were used for each training data set.
Figure~\ref{F:flowchart} shows a diagram illustrating this process.

A training data set informed the statistical model, which included a set of IR absorbance spectra and RON.
Figure~\ref{F:ATR_ex} shows a representative subset of the collected ATR-IR absorbance spectra used with the training data set; shown are isooctane and \emph{n}-heptane neat and as mixtures (PRFX).
The training data set limits the predictive capabilities of the model: the larger and more varied it is, typically the better the model.
The final result is a regression equation, which uses IR absorption spectra as an input and returns a prediction for RON.

To validate the created statistical models, RON was predicted for gasoline samples with known RON.
The fuels studied are the 10 FACE gasolines and 12 FACE gasoline mixtures blended with ethanol~\cite{Cannella:2014aa}.
These include 22 well-documented fuels statistically designed with chemical and ASTM performance variations for researchers to investigate in advanced internal combustion engines.
This bounded the task of creating a model to predict real gasoline fuels that contain hundreds of various hydrocarbons.

%%%%%%%%%%%%%%%%%%%%%%%%%%%%%%%%%%%%%%%%%%%%%%%%%%%%%%%%%%%%%%%%%%%%%%%%%%%%%%%%%%%%%%%%%%%%%%%%%%%%
\section{Results and discussion}
\label{S:results}
%%%%%%%%%%%%%%%%%%%%%%%%%%%%%%%%%%%%%%%%%%%%%%%%%%%%%%%%%%%%%%%%%%%%%%%%%%%%%%%%%%%%%%%%%%%%%%%%%%%%

In order to test how subsets of the fuels considered in this work affect the model predictive performance---i.e., the RON prediction of FACE gasolines, not that of the fuels used to inform the model---we selectively and additively included neat hydrocarbons and surrogate fuels in the model and observed the effect on the residual (residual = actual RON $-$ predicted RON, where the average, max, and min were analyzed).
A limited sensitivity analysis with the neat hydrocarbons was performed first by selectively including them in the model.

The sensitivity study first informed a model with a baseline data set consisting of the six primary neat hydrocarbons and the 134 mixtures, then a prediction of RON for all FACE gasolines.
Next, we selectively included one new neat hydrocarbon in the model (i.e., baseline fuels plus one hydrocarbon) from the additional neat hydrocarbons and observed how the predicted RON of the FACE gasolines were affected.
Here it was learned definitively that the branched alkane and aromatic classes were not satisfactorily represented by isooctane and toluene.
When the model was informed with more neat hydrocarbons representing these classes, the predictive performance changed by up to 40 RON for many of the FACE gasolines.
When including alkanes or aromatics, the average residual improved by 9.0 and 9.7 RON units, respectively.
However, the alkene and cycloalkane classes were sufficiently represented by 1-hexene and methylcyclohexane, respectively; species in these classes respectively improved the residual by 0.4 and 0.7 on average.

\begin{figure}[htbp]
    \centering
    \includegraphics[width=0.7\textwidth]{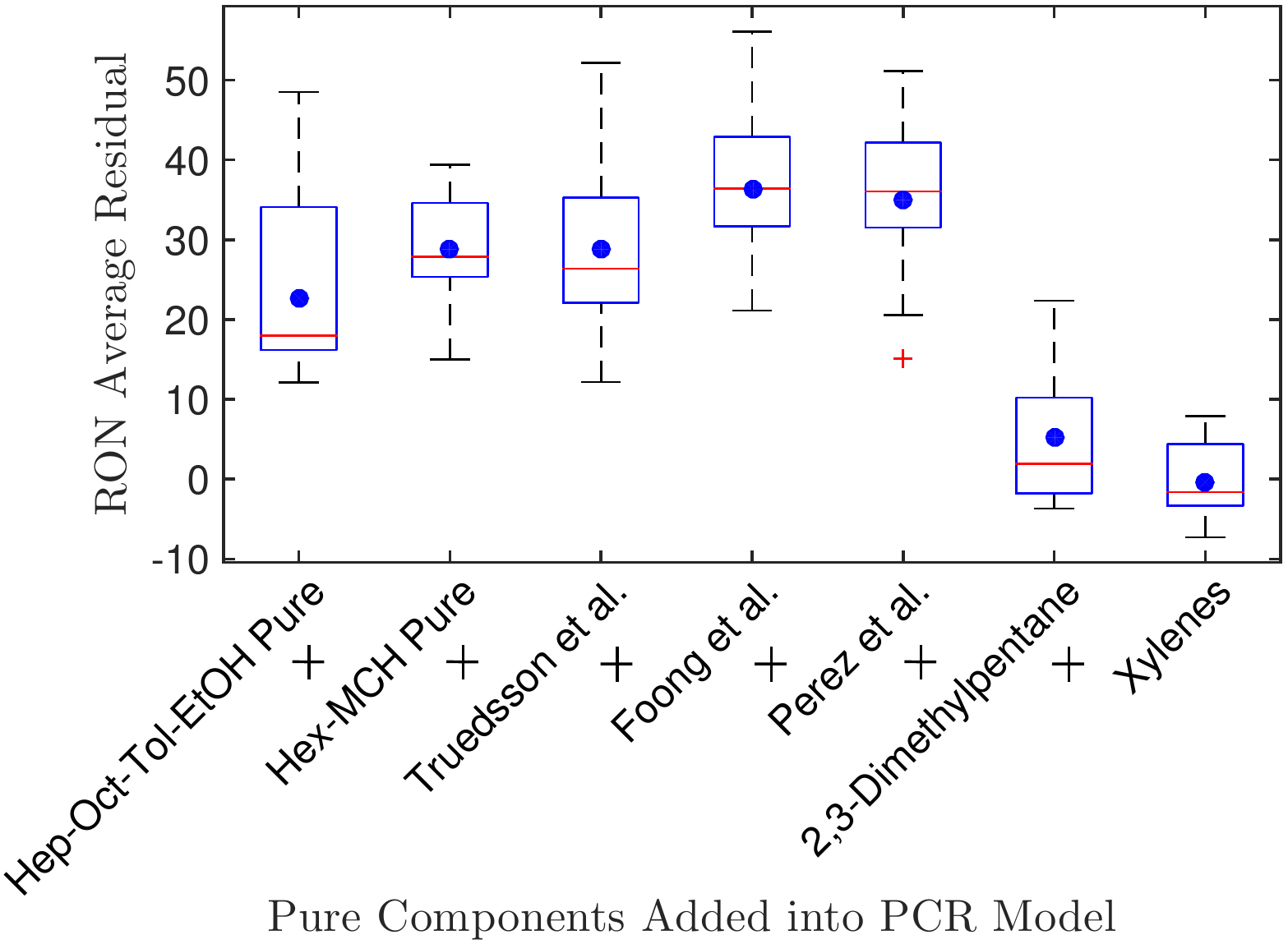}
    \caption{Model performance indicated by average residual for prediction of FACE gasoline RONs as a function of training data set. For each data set, the middle line (red) indicates the median, the circle is the mean, the edges of the box the 25th and 75th percentiles, the whiskers extend to minimum and maximum values not considered outliers, and the outliers are indicated with plus signs. ``Pure'' indicates that the spectra of the pure individual components were used, and ``Hep,'' ``Oct,'' ``Tol,'' ``EtOH,'' ``Hex,'' and ``MCH'' refer to \emph{n}-heptane, isooctane, toluene, ethanol, 1-hexene, and methylcyclohexane, respectively.}
    \label{F:model_improvement}
\end{figure}

Following the sensitivity study, fuels were additively included to inform the model.  
Figure~\ref{F:model_improvement} shows the performance of the model as a function of fuels considered in the training data set.
Moving along the horizontal axis indicates fuels (or fuel sets) additively included in the model (e.g., at the third horizontal axis  location, the model includes six neat hydrocarbons plus the Truedsson et al.~\cite{Truedsson:2014ke} n-heptane, isooctane, toluene, and ethanol fuel blends).  
The vertical axis indicates the RON residual, where a box-and-whisker plot shows the distribution of RON predicted by each training data set---the magnitudes of the maximum and minimum residuals are indicated with error bars (``whiskers''), with outliers beyond a normal distribution indicated with symbols.

\begin{figure}[htbp]
    \centering
    \includegraphics[width=0.7\textwidth]{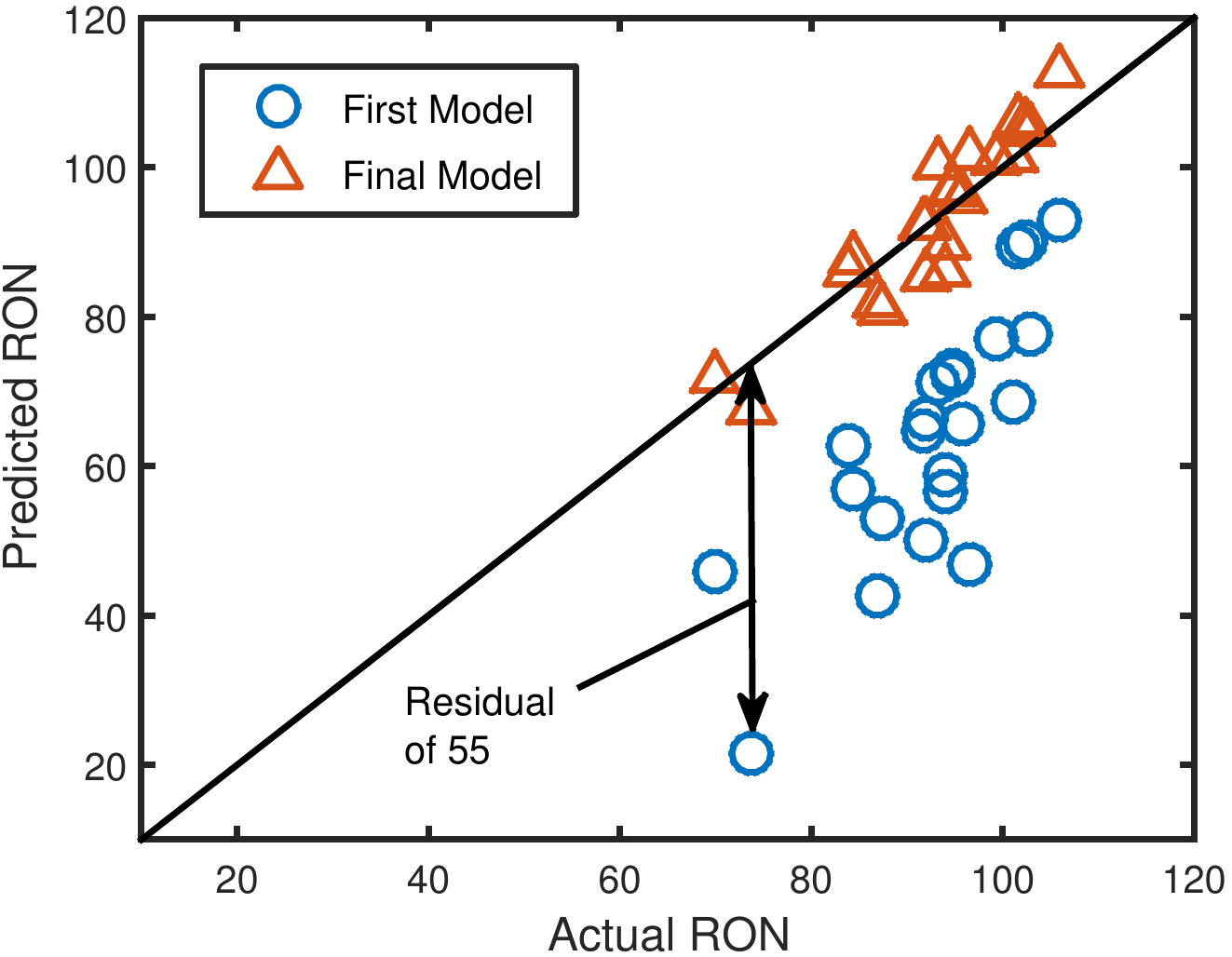}
    \caption{Statistically predicted RON compared with actual RON for the first model (four neat hydrocarbons) and the final model (all neat hydrocarbons and the 134 mixtures considered in this work). Predicted RON values shown are for the FACE gasolines and blends with ethanol. Example residual indicated in figure is actual RON $-$ predicted RON.}
    \label{F:RON_model}
\end{figure}

First, only four neat hydrocarbons were used: \emph{n}-heptane, isooctane, toluene, and ethanol. 
This attempt at modeling resulted in predictions of RON within 22.7\textpm24.8 (residual average \textpm standard deviation), and with an error of 48.5 RON in the worst case.
Next, neat hydrocarbons methylcyclohexane and 1-hexene were added---Perez et al.~\cite{Perez:2012dga} considered these as components in fuel surrogates---which resulted in prediction within 28.8\textpm29.4 RON, with a worst-case error of 39.5 RON.
Following this, we investigated the effect of adding mixtures to the model performance, meaning the absorption spectra and published RONs from literature~\cite{Perez:2012dga,Foong:2014fz,Truedsson:2014ke}.  
We learned that mixtures affected model predictions of RON for many of the FACE gasolines.
Residuals still reached 50 for many of the fuels.
Subsequently, two additional neat hydrocarbons were added---indicated by the sensitivity study to have the largest impact on model performance---representing the branched alkane (2,3-dimethylpentane) and aromatic (xylenes) classes.
The performance of the RON model for all research gasolines converged with the addition of these, resulting in predictions within 0.3\textpm4.4 RON, with a maximum error of 7.9.
Adding additional neat hydrocarbons---whether they represent alkanes, aromatics, cycloalkanes, or alkenes---did not improve the model further.
This indicates that only a few hydrocarbons representing the branched alkane and aromatic classes are required to improve the model, and further inclusion of hydrocarbons resulted in diminishing returns with little effect.
All the additional neat hydrocarbons (see Table~\ref{T:pure_fuels}) were included in the model yielding 0.1\textpm4.8 and 9.7 RON in the worst case.  
Figure~\ref{F:RON_model} illustrates the first (only four of the primary neat hydrocarbons) and the final model performance (all six primary neat hydrocarbons, 28 additional neat hydrocarbons, and the 134 mixtures considered).

\begin{figure}[htbp]
\centering
\begin{subfigure}[b]{0.75\textwidth}
    \includegraphics[width=0.98\linewidth]{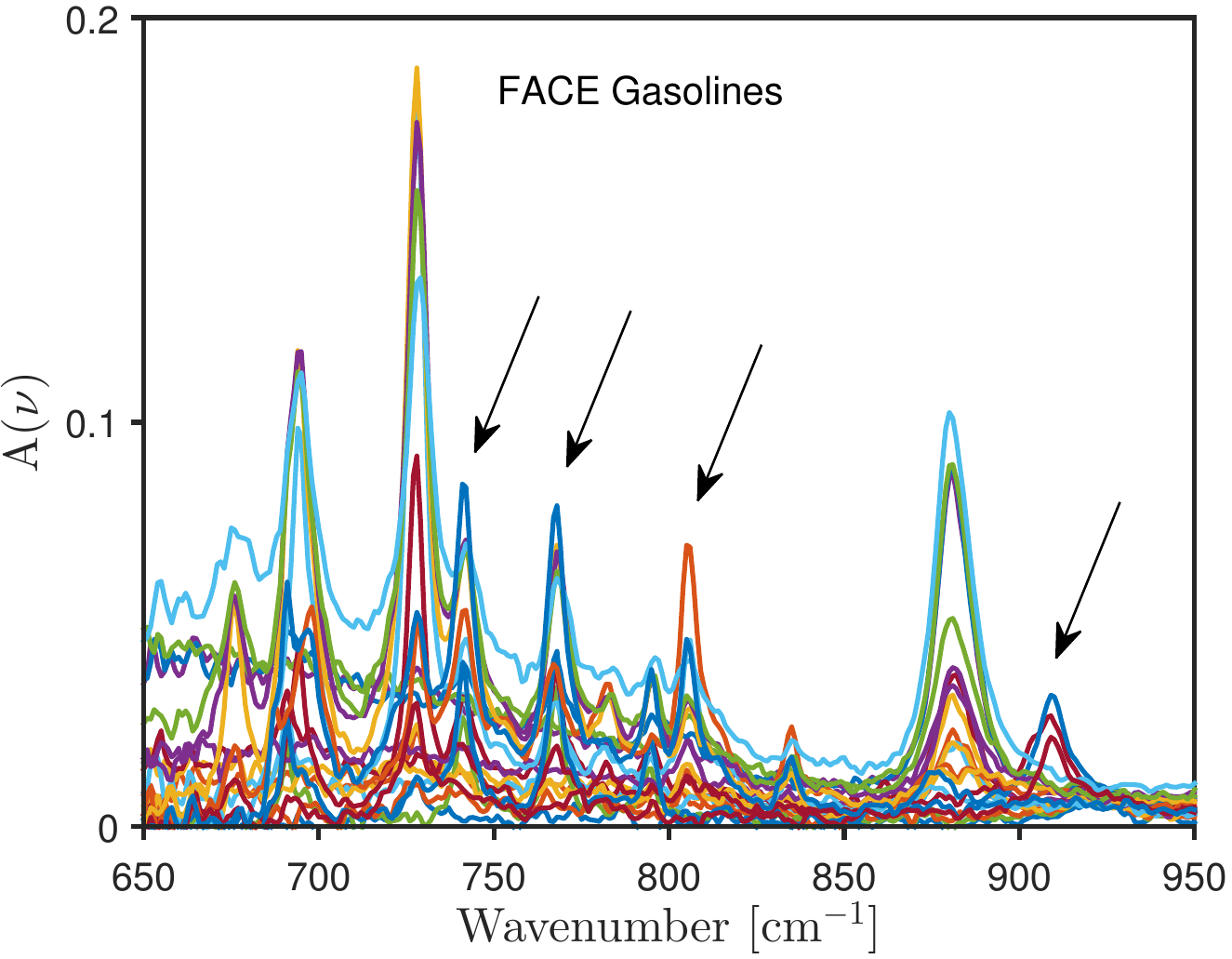}
    \caption{}
    \label{F:FACE_spectra}
\end{subfigure}
\begin{subfigure}[b]{0.75\textwidth}
    \includegraphics[width=\linewidth]{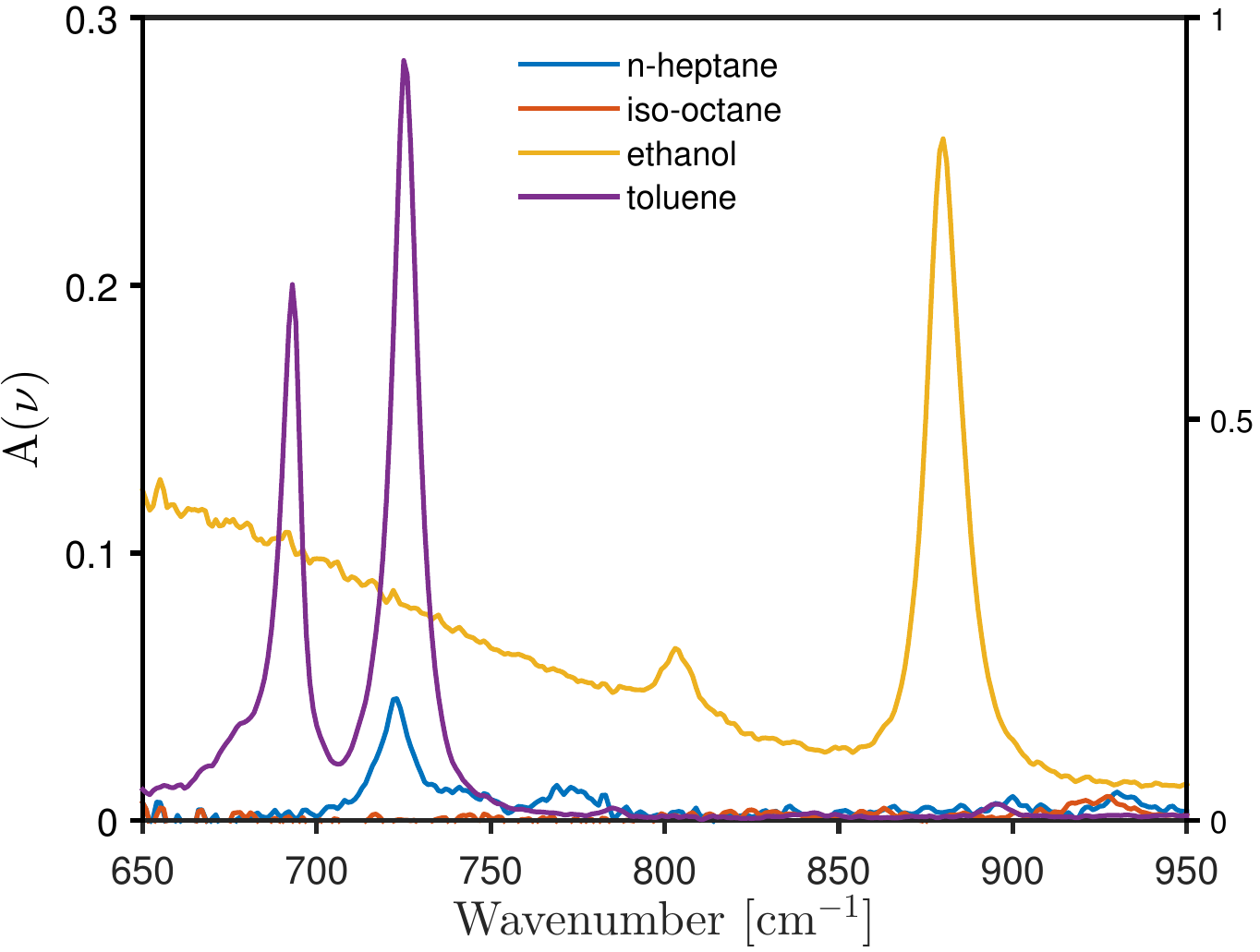}
    \caption{}
    \label{F:Pure_spectra}
\end{subfigure}
\caption{(a) FTIR-ATR absorption spectra for the 10 FACE gasolines and 12 FACE gasoline mixtures blended with ethanol, and (b) initial set of four neat hydrocarbons. Arrows in (a) indicate absorbing frequencies not found in the pure components shown. Right vertical axis in (b) indicates toluene absorbance.}
\end{figure}

The first training data set (recall this included four of the primary neat hydrocarbons) is believed to poorly inform the model as they do not represent the many spectral attributes seen in the FACE gasolines.
Figure~\ref{F:Pure_spectra} depicts FTIR-ATR absorbance spectra (\SIrange{650}{950}{\centi\meter^{-1}}) of these four neat hydrocarbons and the FACE gasolines.
It is evident that additional functional groups need to be included to better represent the FACE gasolines spectroscopically.
In particular, none of the neat hydrocarbons exhibit the absorption peaks of the FACE gasolines at 741, 768, 805, and \SI{909}{\centi\meter^{-1}} (shown by the arrows in Figure~\ref{F:FACE_spectra}).

\begin{figure}[htbp]
    \centering
    \includegraphics[width=0.75\textwidth]{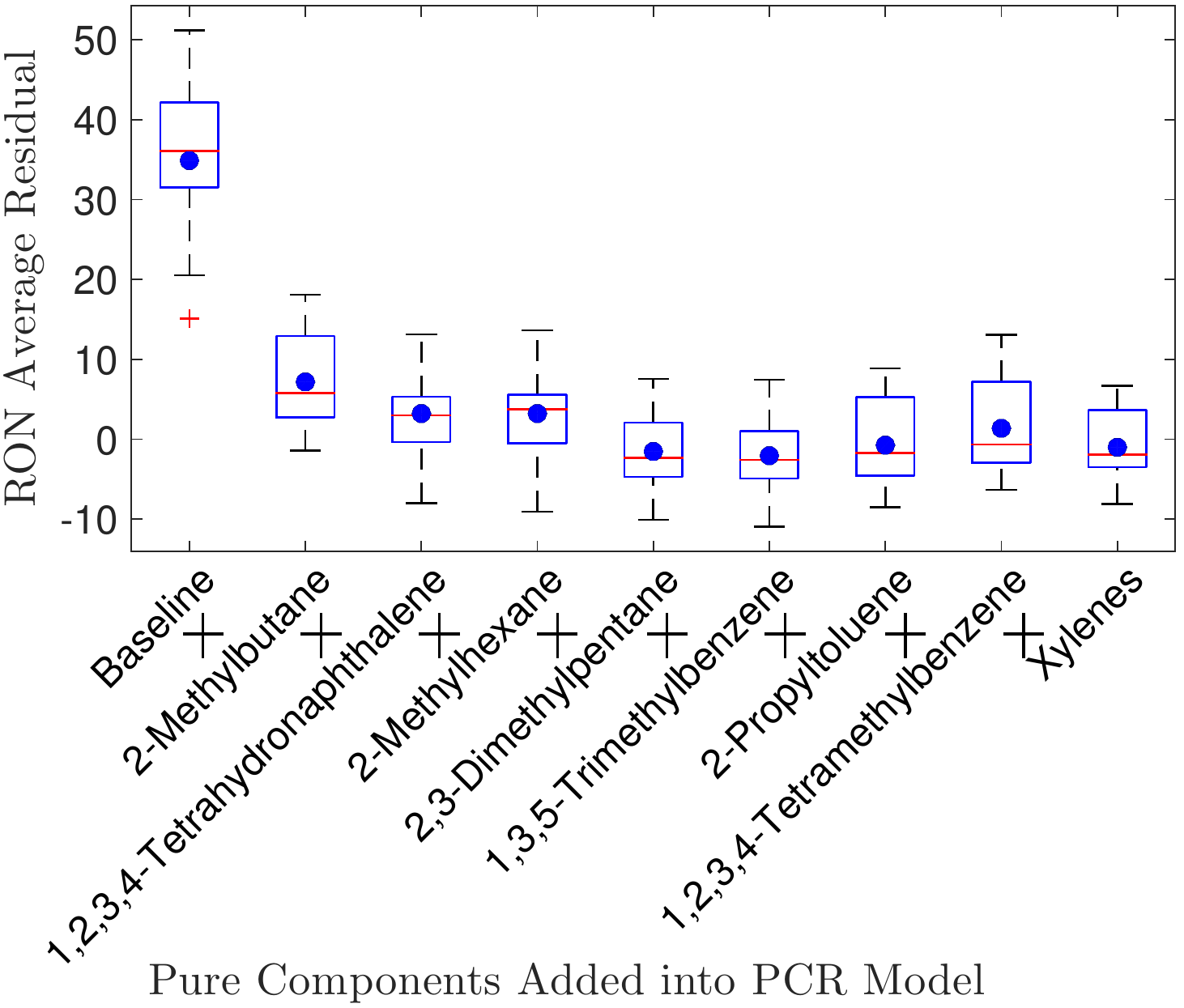}
    \caption{RON residual for all FACE gasolines and blends as a function of neat hydrocarbons added into the model (in a different order than in Fig.~\ref{F:model_improvement}. ``Baseline'' indicates all species added in Fig.~\ref{F:model_improvement} up to the Perez et al.~\cite{Perez:2012dga} fuels. For each dataset, red line indicates the median, blue circle the mean, the edges of the box the 25th and 75th percentiles, the whiskers extend to minimum and maximum values not considered outliers, and the outliers are indicated with red plus signs.}
    \label{F:model_improvement_pure}
\end{figure}

The effect of changing the order in which hydrocarbons were additively included in the model was also investigated, as seen in Figure~\ref{F:model_improvement_pure}.
For example, 2-methylbutane was investigated as a candidate to replace 2,3-dimethylpentane---recall this was originally added as the \emph{first} neat hydrocarbon to the study in Fig.~\ref{F:model_improvement} after the baseline (which yielded 5.1\textpm9.5, max residual = 22.3).
2-Methylbutane improves the model (7.2\textpm9.6, max residual = 18.1) in nearly the same way that 2,3-dimethylpentane originally had.
Following that, we added 1,2,3,4-tetrahydronaphthalene to the model guided by the sensitivity study---recall that this highlighted hydrocarbons with the most impact on model performance.
This hydrocarbon represents the aromatics class, and its addition results in improved model performance (3.1\textpm5.6, max residual = 13.1).
However, this addition is not as significant as that provided by xylene, the second pure component added to the model (0.3\textpm4.4, max residual = 7.9).
Moving to the right in Fig.~\ref{F:model_improvement_pure}, the addition of 2,3-dimethylpentane or xylene no longer dramatically affects the model performance when added after other hydrocarbons; albeit, the model does improve with its addition.

Here we demonstrated that adding pure components from one class can affect the model sensitivity to hydrocarbons from others, and that alternative hydrocarbons representing the branched alkane and aromatic class may be used in lieu of the original 2,3-dimenthylpentane and xylenes considered.
This could be due to shared IR absorbance features at particular optical frequencies.
For example, the fundamental \ce{C-H} stretch frequency from one class overlapping with hydrocarbons from other classes.
This results in spectroscopically redundant information being added to the model, and may explain why some hydrocarbons classes can somewhat inform the statistical model in the same way as other hydrocarbons classes.
Further parametric investigation is needed to fully optimize component choice.

We also explored removing the 134 mixtures from the training data set to observe the effect of only leveraging neat hydrocarbons.
The previously determined (best) performance (0.3\textpm4.4, max residual = 7.9 RON) could not be achieved with pure components alone; the maximum residuals reached 25 RON for many of the predicted FACE gasolines.
We believe mixtures are necessary to inform the model due to non-linear blending effects of RON and the IR spectra with mixtures---this may be attributable to solvation effects.  For example, ethanol particularly introduces these non-linear blending effects for RON~\cite{Foong:2014fz} and the absorbance spectra~\cite{Corsetti2015}.  In general, it is suggested that alcohols interact with hydrocarbons in various ways by means of van der Walls forces, and the molecular structures formed (e.g., double-bonded dimer, linear polymer, water-like structure, etc.) are a function of the alcohol concentration. These various interactions thereby alter the original molecular structure of the hydrocarbons and therefore their absorbing characteristics~\cite{VanNess1967,Reilly2013}.
This highlighted that mixtures are equally as important as the neat components alone for model robustness.
This may indicate the possibility of increasing model performance further by including mixtures containing additional branched alkane and aromatics which improved the model alone.

%%%%%%%%%%%%%%%%%%%%%%%%%%%%%%%%%%%%%%%%%%%%%%%%%%%%%%%%%%%%%%%%%%%%%%%%%%%%%%%%%%%%%%%%%%%%%%%%%%%%%
\section{Conclusions}
\label{S:conclusions}
%%%%%%%%%%%%%%%%%%%%%%%%%%%%%%%%%%%%%%%%%%%%%%%%%%%%%%%%%%%%%%%%%%%%%%%%%%%%%%%%%%%%%%%%%%%%%%%%%%%%%

The approaches of Kelly et al.~\cite{Kelly:1989ui} and other work in this area~\cite{Williams:1990aa,Cooper:1995aa,LitaniBarzilai:1997wp,Korolev:2000tb,Kardamakis:2010db,Swarin:1991aa,Choquette:1996aa,Fodor:1996aa,Balabin:2008aa,Monteiro:2009kk,Morris:2009cg,Veras:2010jt,Tomren:2012aa} used real-world fuel samples to inform the optics-based statistical models.
This work instead used neat hydrocarbons---six of them are primarily utilized as constituents in gasoline surrogates, and 28 being the primary constituents in the FACE gasolines---as well as mixtures that contain the primary six components.
The six primary pure components and 134 mixtures of these pure components predicted RON poorly (34.8\textpm36.1 on average and 51.2 RON in the worst case).
However, the addition of two neat hydrocarbons, one to each represent the branched alkane and aromatic classes, resulted in model improvement: predicted RON within 0.3\textpm4.4 and 7.9 RON in the worst case, respectively.
This performance could be achieved with various neat hydrocarbons representing these classes.
Additional parametric investigation is required for ideal fuel selection.
This, however, would be difficult to determine due to the many possible combinations and results to likely change when considering additional hydrocarbons.   
More importantly and simpler to implement, mixtures containing these two additional pure components should be investigated for their effect on the model, primarily because the six primary neat hydrocarbons proved to be important in mixtures as they are individually.

This work shows that the ignition quality of gasolines can be represented by as few as eight hydrocarbon spectra.
With this information, the most impactful fuels (neat or otherwise) can be targeted to inform spectroscopic surrogates to predict performance of complex fuels, in this case the FACE gasolines.
We developed a model informed by simple fuel surrogates and neat hydrocarbons to predict RON of the FACE gasolines.
Therefore, this work builds upon previous efforts by creating models that are irrespective of the fuels we wish to predict performance attributes of.
The primary benefit being predicting fuel performance without the need to gather a training data set that uses those particular fuels to inform the model---e.g., using ATR-FTIR spectra and known RONs of characterized gasolines to inform the model to predict RON for an unknown gasoline sample from its measured ATR-FTIR spectra.
In addition, the results may support using computationally determined performance metrics for the training data set.
This is to say, fuel performance metrics of real fuels---burdensome to accurately model due to complex chemical mechanisms---can be predicted with a computationally modeled data set of neat-hydrocarbons and surrogate fuels, which are relatively simple and computationally more efficient to model.
This model would predict real fuel performance (e.g., RON or alternative metrics) informed by computational simulations and FTIR-ATR absorption spectroscopy.

%%%%%%%%%%%%%%%%%%%%%%%%%%%%%%%%%%%%%%%%%%%%%%%%%%%%%%%%%%%%%%%%%%%%%%%%%%%%%%%%%%%%%%%%%%%%%%%%%%%
\section*{Acknowledgments}
%%%%%%%%%%%%%%%%%%%%%%%%%%%%%%%%%%%%%%%%%%%%%%%%%%%%%%%%%%%%%%%%%%%%%%%%%%%%%%%%%%%%%%%%%%%%%%%%%%%

The authors gratefully acknowledge the Chevron Energy Technology Company for supporting this research, and Dr. Jeffrey Gautschi for kind access to his lab and FTIR equipment.

%%%%%%%%%%%%%%%%%%%%%%%%%%%%%%%%%%%%%%%%%%%%%%%%%%%%%%%%%%%%%%%%%%%%%%%%%%%%%%%%%%%%%%%%%%%%%%%%%%%
%% The Appendices part starts with the command \appendix;

%%%%%%%%%%%%%%%%%%%%%%%%%%%%%%%%%%%%%%%%%%%%%%%%%%%%%%%%%%%%%%%%%%%%%
%% Place the appropriate \bibliography command here.

\clearpage

\bibliography{refs}
\bibliographystyle{elsarticle-num}

\end{document}